\documentclass[aps,prl,twocolumn,showpacs,superscriptaddress,nofootinbib,floatfix]{revtex4-1}

\usepackage{amsmath}   
\usepackage{graphicx}  
\graphicspath{{figures/}}
\usepackage{xspace}
\usepackage{units}
\usepackage{accents}
\usepackage{scalerel}
\usepackage{color}
\usepackage{bm}
\usepackage{subfigure}
\usepackage{alphalph}

\usepackage{hyperref}

\newlength{\heightnu}
\AtBeginDocument{\settoheight{\heightnu}{$\nu$}}

\newif\ifpdf
\ifx\pdfoutput\undefined
   \pdffalse
\else
   \pdfoutput=1
   \pdftrue
\fi
\ifpdf
   \usepackage{graphicx}
   \usepackage{epstopdf}
\else
   \usepackage{graphicx}
\fi

\begin{document}

\title{Constraint on the Matter-Antimatter Symmetry-Violating Phase in Neutrino Oscillations}




\newcommand{\INSTHD}{\affiliation{University Autonoma Madrid, Department of Theoretical Physics, 28049 Madrid, Spain}}
\newcommand{\INSTEE}{\affiliation{University of Bern, Albert Einstein Center for Fundamental Physics, Laboratory for High Energy Physics (LHEP), Bern, Switzerland}}
\newcommand{\INSTFE}{\affiliation{Boston University, Department of Physics, Boston, Massachusetts, U.S.A.}}
\newcommand{\INSTD}{\affiliation{University of British Columbia, Department of Physics and Astronomy, Vancouver, British Columbia, Canada}}
\newcommand{\INSTGA}{\affiliation{University of California, Irvine, Department of Physics and Astronomy, Irvine, California, U.S.A.}}
\newcommand{\INSTI}{\affiliation{IRFU, CEA Saclay, Gif-sur-Yvette, France}}
\newcommand{\INSTGB}{\affiliation{University of Colorado at Boulder, Department of Physics, Boulder, Colorado, U.S.A.}}
\newcommand{\INSTFG}{\affiliation{Colorado State University, Department of Physics, Fort Collins, Colorado, U.S.A.}}
\newcommand{\INSTFH}{\affiliation{Duke University, Department of Physics, Durham, North Carolina, U.S.A.}}
\newcommand{\INSTBA}{\affiliation{Ecole Polytechnique, IN2P3-CNRS, Laboratoire Leprince-Ringuet, Palaiseau, France }}
\newcommand{\INSTEF}{\affiliation{ETH Zurich, Institute for Particle Physics and Astrophysics, Zurich, Switzerland}}
\newcommand{\INSTIE}{\affiliation{CERN European Organization for Nuclear Research, CH-1211 Genève 23, Switzerland}}
\newcommand{\INSTEG}{\affiliation{University of Geneva, Section de Physique, DPNC, Geneva, Switzerland}}
\newcommand{\INSTHJ}{\affiliation{University of Glasgow, School of Physics and Astronomy, Glasgow, United Kingdom}}
\newcommand{\INSTDG}{\affiliation{H. Niewodniczanski Institute of Nuclear Physics PAN, Cracow, Poland}}
\newcommand{\INSTCB}{\affiliation{High Energy Accelerator Research Organization (KEK), Tsukuba, Ibaraki, Japan}}
\newcommand{\INSTIB}{\affiliation{University of Houston, Department of Physics, Houston, Texas, U.S.A.}}
\newcommand{\INSTED}{\affiliation{Institut de Fisica d'Altes Energies (IFAE), The Barcelona Institute of Science and Technology, Campus UAB, Bellaterra (Barcelona) Spain}}
\newcommand{\INSTEC}{\affiliation{IFIC (CSIC \& University of Valencia), Valencia, Spain}}
\newcommand{\INSTHH}{\affiliation{Institute For Interdisciplinary Research in Science and Education (IFIRSE), ICISE, Quy Nhon, Vietnam}}
\newcommand{\INSTEI}{\affiliation{Imperial College London, Department of Physics, London, United Kingdom}}
\newcommand{\INSTGF}{\affiliation{INFN Sezione di Bari and Universit\`a e Politecnico di Bari, Dipartimento Interuniversitario di Fisica, Bari, Italy}}
\newcommand{\INSTBE}{\affiliation{INFN Sezione di Napoli and Universit\`a di Napoli, Dipartimento di Fisica, Napoli, Italy}}
\newcommand{\INSTBF}{\affiliation{INFN Sezione di Padova and Universit\`a di Padova, Dipartimento di Fisica, Padova, Italy}}
\newcommand{\INSTBD}{\affiliation{INFN Sezione di Roma and Universit\`a di Roma ``La Sapienza'', Roma, Italy}}
\newcommand{\INSTEB}{\affiliation{Institute for Nuclear Research of the Russian Academy of Sciences, Moscow, Russia}}
\newcommand{\INSTHI}{\affiliation{International Centre of Physics, Institute of Physics (IOP), Vietnam Academy of Science and Technology (VAST), 10 Dao Tan, Ba Dinh, Hanoi, Vietnam}}
\newcommand{\INSTHA}{\affiliation{Kavli Institute for the Physics and Mathematics of the Universe (WPI), The University of Tokyo Institutes for Advanced Study, University of Tokyo, Kashiwa, Chiba, Japan}}
\newcommand{\INSTID}{\affiliation{Keio University, Department of Physics, Kanagawa, Japan}}
\newcommand{\INSTIF}{\affiliation{King's College London, Department of Physics, Strand, London WC2R 2LS, United Kingdom}}
\newcommand{\INSTCC}{\affiliation{Kobe University, Kobe, Japan}}
\newcommand{\INSTCD}{\affiliation{Kyoto University, Department of Physics, Kyoto, Japan}}
\newcommand{\INSTEJ}{\affiliation{Lancaster University, Physics Department, Lancaster, United Kingdom}}
\newcommand{\INSTFC}{\affiliation{University of Liverpool, Department of Physics, Liverpool, United Kingdom}}
\newcommand{\INSTFI}{\affiliation{Louisiana State University, Department of Physics and Astronomy, Baton Rouge, Louisiana, U.S.A.}}
\newcommand{\INSTHB}{\affiliation{Michigan State University, Department of Physics and Astronomy,  East Lansing, Michigan, U.S.A.}}
\newcommand{\INSTCE}{\affiliation{Miyagi University of Education, Department of Physics, Sendai, Japan}}
\newcommand{\INSTDF}{\affiliation{National Centre for Nuclear Research, Warsaw, Poland}}
\newcommand{\INSTFJ}{\affiliation{State University of New York at Stony Brook, Department of Physics and Astronomy, Stony Brook, New York, U.S.A.}}
\newcommand{\INSTGJ}{\affiliation{Okayama University, Department of Physics, Okayama, Japan}}
\newcommand{\INSTCF}{\affiliation{Osaka City University, Department of Physics, Osaka, Japan}}
\newcommand{\INSTGG}{\affiliation{Oxford University, Department of Physics, Oxford, United Kingdom}}
\newcommand{\INSTGC}{\affiliation{University of Pittsburgh, Department of Physics and Astronomy, Pittsburgh, Pennsylvania, U.S.A.}}
\newcommand{\INSTFA}{\affiliation{Queen Mary University of London, School of Physics and Astronomy, London, United Kingdom}}
\newcommand{\INSTE}{\affiliation{University of Regina, Department of Physics, Regina, Saskatchewan, Canada}}
\newcommand{\INSTGD}{\affiliation{University of Rochester, Department of Physics and Astronomy, Rochester, New York, U.S.A.}}
\newcommand{\INSTHC}{\affiliation{Royal Holloway University of London, Department of Physics, Egham, Surrey, United Kingdom}}
\newcommand{\INSTBC}{\affiliation{RWTH Aachen University, III. Physikalisches Institut, Aachen, Germany}}
\newcommand{\INSTFB}{\affiliation{University of Sheffield, Department of Physics and Astronomy, Sheffield, United Kingdom}}
\newcommand{\INSTDI}{\affiliation{University of Silesia, Institute of Physics, Katowice, Poland}}
\newcommand{\INSTIA}{\affiliation{SLAC National Accelerator Laboratory, Stanford University, Menlo Park, California, USA}}
\newcommand{\INSTBB}{\affiliation{Sorbonne Universit\'e, Universit\'e Paris Diderot, CNRS/IN2P3, Laboratoire de Physique Nucl\'eaire et de Hautes Energies (LPNHE), Paris, France}}
\newcommand{\INSTEH}{\affiliation{STFC, Rutherford Appleton Laboratory, Harwell Oxford,  and  Daresbury Laboratory, Warrington, United Kingdom}}
\newcommand{\INSTCH}{\affiliation{University of Tokyo, Department of Physics, Tokyo, Japan}}
\newcommand{\INSTBJ}{\affiliation{University of Tokyo, Institute for Cosmic Ray Research, Kamioka Observatory, Kamioka, Japan}}
\newcommand{\INSTCG}{\affiliation{University of Tokyo, Institute for Cosmic Ray Research, Research Center for Cosmic Neutrinos, Kashiwa, Japan}}
\newcommand{\INSTHF}{\affiliation{Tokyo Institute of Technology, Department of Physics, Tokyo, Japan}}
\newcommand{\INSTGI}{\affiliation{Tokyo Metropolitan University, Department of Physics, Tokyo, Japan}}
\newcommand{\INSTHG}{\affiliation{Tokyo University of Science, Faculty of Science and Technology, Department of Physics, Noda, Chiba, Japan}}
\newcommand{\INSTF}{\affiliation{University of Toronto, Department of Physics, Toronto, Ontario, Canada}}
\newcommand{\INSTB}{\affiliation{TRIUMF, Vancouver, British Columbia, Canada}}
\newcommand{\INSTG}{\affiliation{University of Victoria, Department of Physics and Astronomy, Victoria, British Columbia, Canada}}
\newcommand{\INSTDJ}{\affiliation{University of Warsaw, Faculty of Physics, Warsaw, Poland}}
\newcommand{\INSTDH}{\affiliation{Warsaw University of Technology, Institute of Radioelectronics and Multimedia Technology, Warsaw, Poland}}
\newcommand{\INSTFD}{\affiliation{University of Warwick, Department of Physics, Coventry, United Kingdom}}
\newcommand{\INSTGH}{\affiliation{University of Winnipeg, Department of Physics, Winnipeg, Manitoba, Canada}}
\newcommand{\INSTEA}{\affiliation{Wroclaw University, Faculty of Physics and Astronomy, Wroclaw, Poland}}
\newcommand{\INSTHE}{\affiliation{Yokohama National University, Faculty of Engineering, Yokohama, Japan}}
\newcommand{\INSTH}{\affiliation{York University, Department of Physics and Astronomy, Toronto, Ontario, Canada}}

\INSTHD
\INSTEE
\INSTFE
\INSTD
\INSTGA
\INSTI
\INSTGB
\INSTFG
\INSTFH
\INSTBA
\INSTEF
\INSTIE
\INSTEG
\INSTHJ
\INSTDG
\INSTCB
\INSTIB
\INSTED
\INSTEC
\INSTHH
\INSTEI
\INSTGF
\INSTBE
\INSTBF
\INSTBD
\INSTEB
\INSTHI
\INSTHA
\INSTID
\INSTIF
\INSTCC
\INSTCD
\INSTEJ
\INSTFC
\INSTFI
\INSTHB
\INSTCE
\INSTDF
\INSTFJ
\INSTGJ
\INSTCF
\INSTGG
\INSTGC
\INSTFA
\INSTE
\INSTGD
\INSTHC
\INSTBC
\INSTFB
\INSTDI
\INSTIA
\INSTBB
\INSTEH
\INSTCH
\INSTBJ
\INSTCG
\INSTHF
\INSTGI
\INSTHG
\INSTF
\INSTB
\INSTG
\INSTDJ
\INSTDH
\INSTFD
\INSTGH
\INSTEA
\INSTHE
\INSTH

\author{K.\,Abe}\INSTBJ
\author{R.\,Akutsu}\INSTCG
\author{A.\,Ali}\INSTCD
\author{C.\,Alt}\INSTEF
\author{C.\,Andreopoulos}\INSTEH\INSTFC
\author{L.\,Anthony}\INSTFC
\author{M.\,Antonova}\INSTEC
\author{S.\,Aoki}\INSTCC
\author{A.\,Ariga}\INSTEE
\author{Y.\,Asada}\INSTHE
% ...

\collaboration{The T2K Collaboration}\noaffiliation

\date{\today}

\maketitle

\clearpage

\section{Introduction}
The symmetry between matter and antimatter, charge-conjugation parity-reversal ($CP$) symmetry, was believed to be a perfect symmetry of nature. In 1964, asymmetric behavior of particles and anti-particles, $CP$ violation, was observed~\cite{Christenson:1964fg}, and $CP$ violation in the weak interactions of quarks was soon established~\cite{Tanabashi:2018oca}. Sakharov proposed~\cite{Sakharov:1967dj} that $CP$ violation is one of the necessary conditions for an explanation of the observed imbalance of matter and antimatter abundance in the universe. However, the $CP$ violation in quarks is too small to provide this explanation. To date, we have not found $CP$ violation in any non-quark elementary particle systems. It has been shown that $CP$ violation in the lepton sector could generate the matter-antimatter disparity through the process called leptogenesis~\cite{Fukugita:1986hr}. Leptonic mixing, which appears in the Standard Model charged current interactions~\cite{Fukuda:1998mi,Ahmad:2002jz}, provides a potential source of $CP$ violation through a complex phase $\delta_{CP}$, which may have consequences for theoretical models of leptogenesis~\cite{Pascoli:2006ie,Hagedorn:2017wjy,RevModPhys.84.515}. This $CP$ violation can be measured in muon neutrino to electron neutrino oscillations and the corresponding antineutrino oscillations, which are experimentally accessible with accelerator-produced beams as established by the Tokai-to-Kamioka (T2K) and NOvA experiments~\cite{Abe:2013hdq,Acero:2019ksn}.  Until now, the value of $\delta_{CP}$ has not been significantly constrained by neutrino oscillation experiments. Here the T2K collaboration reports a measurement that favors large enhancement of the neutrino oscillation probability, excluding values of $\delta_{CP}$ which result in a large enhancement of the observed antineutrino oscillation probability at three standard deviations ($3\sigma$). The $3\sigma$ confidence level interval for $\delta_{CP}$, which is cyclic and repeats every $2\pi$, is [$-3.41$,$-0.03$] for the so-called normal mass ordering, and [$-2.54$,$-0.32$] for the inverted mass ordering. Our results show an indication of $CP$ violation in the lepton sector. Herein we establish methods for sensitive searches for a matter-antimatter asymmetry in neutrino oscillations using accelerator-produced neutrino beams. Future measurements with larger data samples will test whether leptonic $CP$ violation is larger than the $CP$ violation in the quark sector.

\section{Main}
Previous observations of neutrino oscillations have established that the three known neutrino flavour states, $\nu_{e}$, $\nu_{\mu}$ and $\nu_{\tau}$ are mixtures of three mass states, $\nu_{1}$, $\nu_{2}$ and $\nu_{3}$~\cite{Abe:2018wpn,Abe:2008aa,Adey:2018zwh,Aharmim:2011vm}. This mixing is described by a unitary matrix called the Pontecorvo-Maki-Nakagawa-Sakata (PMNS) matrix \cite{Maki:1962mu,Pontecorvo:1967fh}, which can be parameterized by three mixing angles $\theta_{12}$, $\theta_{13}$ and $\theta_{23}$, and complex phases. Of these phases, neutrino oscillations are sensitive to $\delta_{CP}$. The probabilities for the neutrinos to oscillate from one flavour state to another as they travel depend on these mixing parameters and the mass squared differences ($\Delta m^{2}_{ij}=m^{2}_{i}-m^{2}_{j}$) between the neutrino mass states. The PMNS parameters and the mass squared differences are referred to as ``oscillation parameters". It is known that $\nu_{1}$ and $\nu_{2}$ lie close to each other in mass, with $\Delta m^{2}_{21}=(7.53\pm 0.18)\times 10^{-5}$\,$\text{eV}^{2}/c^{4}$, while $\lvert \Delta m^{2}_{32}\lvert$ is approximately 30 times larger. However, it is not known whether $m_{3}$ has a larger or smaller mass than $m_{1}$ and $m_{2}$~\cite{Tanabashi:2018oca}. The case where the mass of $m_{3}$ is larger (smaller) is called the normal (inverted) ordering. The $CP$ symmetry violating effect in neutrino and antineutrino oscillations has a magnitude that depends on the Jarlskog invariant \cite{Krastev:1988yu,Jarlskog:1985cw}:
\begin{equation}
J_{CP,l}=\frac{1}{8}\cos \theta_{13} \sin(2\theta_{12})\sin(2\theta_{23})\sin(2\theta_{13})\sin(\delta_{CP})
\end{equation} . 
According to current measurements, this is approximately $0.033\sin(\delta_{CP})$~\cite{Tanabashi:2018oca}. This value has the potential to be three orders of magnitude larger than the measured quark sector $CP$ violation ($J_{CP,q}=3\times 10^{-5}$)~\cite{Tanabashi:2018oca}. Prior to this work, no experiment has excluded any values of $\delta_{CP}$ (taking into account both mass orderings) at the 99.73\% ($3\sigma$) confidence level, considered as evidence in the particle physics community.

T2K is a long-baseline neutrino experiment that uses beams of muon neutrinos and antineutrinos, with energy spectra peaked at 0.6\,GeV. We observe interactions of the neutrinos at a near detector facility 280\,m from the beam production point which characterizes the beam and the interactions of the neutrinos before oscillations. The beam then propagates 295\,km through the Earth to the T2K far detector, Super-Kamiokande (SK). SK measures the oscillated beam, which gives sensitivity to the oscillation parameters. 

\begin{figure}
\centering
\includegraphics[width=.49\textwidth]{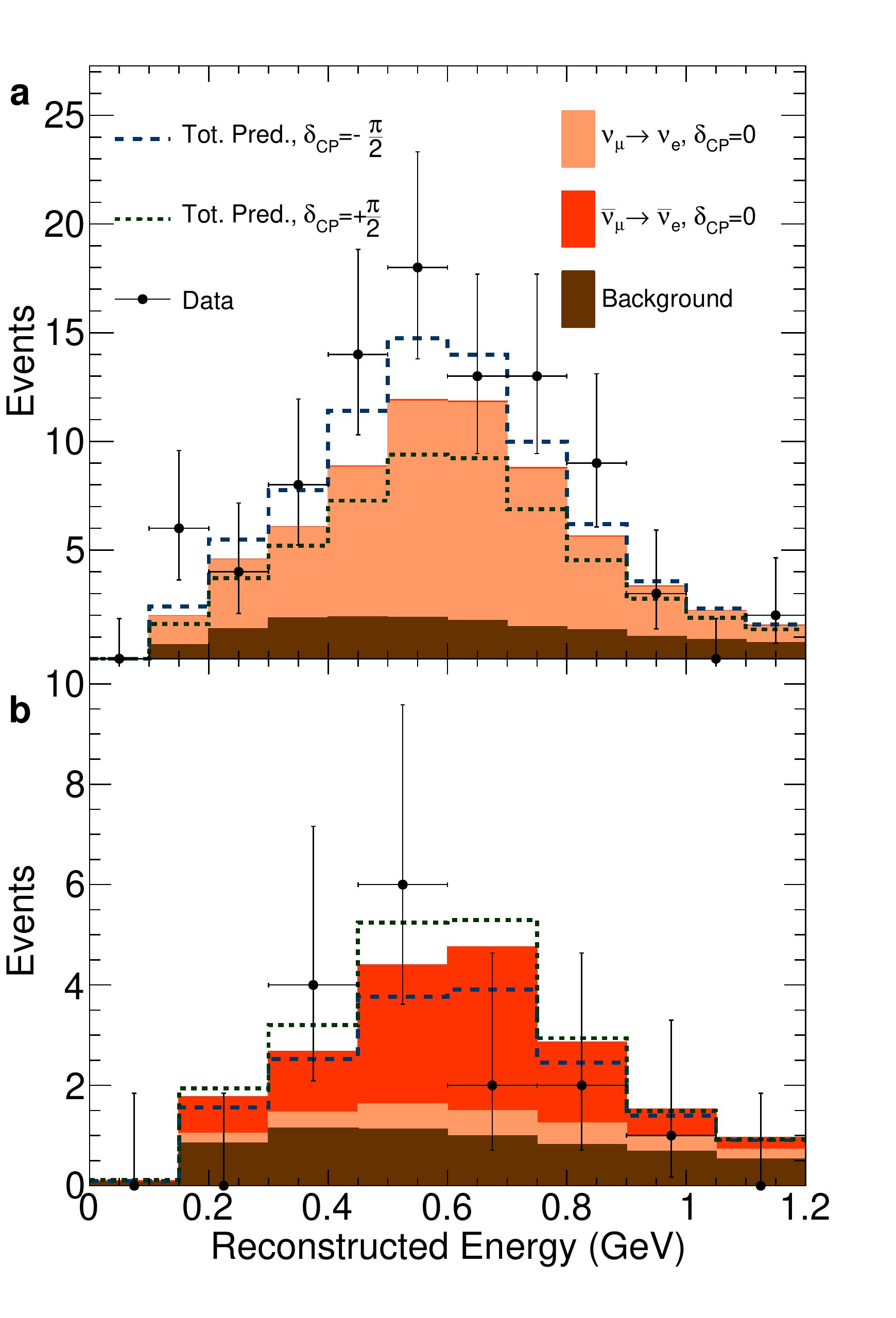}
\begin{tabular}{c|c|c|c}
        {\large \textbf{c}} & 1$e$0de $\nu$-mode & 1$e$0de $\bar{\nu}$-mode & 1$e$1de $\nu$-mode \\
        \hline
        $\nu_{\mu}\rightarrow\nu_{e}$ & 59.0 & 3.0 & 5.4 \\
        $\bar{\nu}_{\mu}\rightarrow\bar{\nu}_{e}$ & 0.4 & 7.5 & 0.0 \\
        Background & 13.8 & 6.4 & 1.5 \\
        \hline
        Total Pred. & 73.2 & 16.9 & 6.9 \\
        Syst. Unc. & 8.8\% & 7.1\% & 18.4\% \\
        \hline
        Data & 75 & 15 & 15 \\
        
    \end{tabular}
\caption{\textbf{Observed $\nu_{e}$ and $\bar{\nu}_{e}$ candidate events at SK}. Subfigure \textbf{a} (\textbf{b}) shows the reconstructed neutrino energy spectra for the SK samples containing electron-like events in neutrino(antineutrino)-mode beam running. The uncertainty shown around the data points accounts for statistical uncertainty. The uncertainty range is chosen to include all points for which the measured number of data events is inside the 68\% confidence interval of a Poisson distribution centred at that point. The solid stacked chart shows the predicted number of events for the $CP$-conserving point $\delta_{CP}=0$ separated according to whether the event was from an oscillated neutrino or antineutrino or from a background process. The dashed lines show the total predicted number of events for the two most extreme $CP$-violating cases. Subfigure \textbf{c} shows the predicted number of events for $\delta_{CP}=-\frac{\pi}{2}$ and the measured number of events in the three electron-like samples at SK. The predicted number of events is broken down into the same categories as subfigures \textbf{a} and \textbf{b} and the systematic uncertainty shown is after the near-detector fit. In both \textbf{a} and \textbf{b} for all predictions, normal ordering is assumed, and $\sin^2\theta_{23}$ and $\Delta m^2_{32}$ are at their best-fit values.  $\sin^2\theta_{13}$, $\sin^2\theta_{12}$ and $\Delta m^2_{21}$ take the values indicated by external world average measurements~\cite{Tanabashi:2018oca}. The parameters accounting for systematic uncertainties take their best-fit values after the near-detector fit.}
\label{fig:nueerecspectra}
\end{figure}

For this beam energy and propagation distance, the probability for muon neutrinos(antineutrinos) to oscillate to electron neutrinos(antineutrinos) is given at leading order in $\delta_{CP}$, including the $CP$-violating term but neglecting effects from propagation through matter, by:
\begin{equation}
   \begin{split}
   \text{P}\left(\nu_{\mu}\rightarrow\nu_{e}\right)\approx\sin^{2}(2\theta_{13})\sin^{2}(\theta_{23})\sin^{2}\left(\frac{1.27\Delta m^2_{32}L}{E}\right)\\
   \mp\frac{1.27\Delta m^2_{21}L}{E}8J_{CP}\sin^2\left(\frac{1.27\Delta m^2_{32}L}{E}\right).    
   \end{split}
   \label{eq:nueprob}
\end{equation}
Here, $E$ is the energy of the neutrino in GeV, the mass squared differences are given in $\text{eV}^{2}/c^{4}$ and $L$ is the propagation baseline in km. The second term in Eq.~\ref{eq:nueprob} has a negative sign for neutrinos and a positive sign for antineutrinos. The baseline and beam energy are optimised so that at our baseline, the probability to oscillate to electron neutrinos reaches a maximum at energies around the beam energy. While the probability of oscillation to electron neutrinos is small, muon neutrinos also oscillate to tau neutrinos, which are not identifiable at SK as T2K's beam energy is too low for a charged tau lepton to be produced. Overall, the probability for muon neutrinos and antineutrinos to maintain their initial flavour is:
\begin{equation}
   \begin{split}
    \text{P}(\nu_{\mu}&\rightarrow\nu_{\mu})\approx1-4\cos^{2}(\theta_{13})\sin^{2}(\theta_{23})\\
    &\times\left[1-\cos^{2}(\theta_{13})\sin^{2}(\theta_{23})\right]\sin^{2}\left(\frac{1.27\Delta m^2_{32}L}{E}\right).
   \end{split}
    \label{eq:numuprob}
\end{equation}
As the probability for oscillation to tau neutrinos is large at our modal beam energy and baseline, there is a minimum in the muon neutrino energy spectrum. 
The position of this minimum gives the experiment sensitivity to the magnitude of $\Delta m^{2}_{32}$ and the depth gives sensitivity to $\sin^{2}(2\theta_{23})$. The height of the peak in the electron neutrino energy spectrum at the oscillation maximum is, at leading order, determined by $\sin^{2}(\theta_{23})$ and $\sin^{2}(2\theta_{13})$ (see Eq.~\ref{eq:nueprob}). However, it also has a sub-leading dependence on $\delta_{CP}$ and the neutrino mass ordering, giving sensitivity to these parameters. Due to this interdependence, determining the other PMNS mixing parameters is important in measuring $\delta_{CP}$. As can be seen from Figure \ref{fig:nueerecspectra}, changing $\delta_{CP}$ from $+\frac{\pi}{2}$ to $-\frac{\pi}{2}$ can lead to $\mathcal{O}$(40\%) changes in the number of electron neutrinos expected at SK. In our analysis we model the observed kinematic distributions of the final state particles using the full oscillation probability including the effect of the neutrinos propagating through matter, which is an O(10\%) perturbation to the probability discussed in Equations~\ref{eq:nueprob} \& \ref{eq:numuprob}~\cite{PhysRevD.22.2718}.

The T2K neutrino beam is generated at the Japan Proton Accelerator Research Complex (J-PARC) by impinging a 30~GeV beam of protons onto a graphite target~\cite{Abe:2011ks}. This interaction creates a large number of secondary hadrons, which are focused using magnetic horns. A neutrino(antineutrino)-enhanced beam is selected by focusing  positively(negatively)-charged particles --dominantly pions--, by choosing the polarity of the magnetic field produced by the horns, thereby enabling us to study the differences between neutrino and antineutrino oscillations. The beam axis is directed 2.5$^{\circ}$ away from the SK detector, taking advantage of the kinematics of the two-body pion decay to produce a narrow neutrino spectrum peaked at the expected energy of maximum oscillation probability~\cite{osti_52878}. The results reported here are based on SK data collected between 2009 and 2018 in neutrino(antineutrino) mode and include a beam exposure of $1.49\times 10^{21}$ ($1.64\times 10^{21}$) protons hitting the T2K neutrino production target. 

Neutrinos are detected by observing the particles they produce when they interact. At neutrino energies of 0.6~GeV the dominant interaction process is Charged-Current Quasi-Elastic (CCQE) scattering via the exchange of a W boson with a single neutron or proton bound in the target nucleus. In this process the neutrino (antineutrino) turns into a charged lepton (antilepton) of the same flavour. We are thereby able to identify the incoming neutrino's flavour. 

 Our near detector facility consists of two detectors both located 280~m downstream of the beam production target~\cite{Abe:2011ks}. The INGRID detector~\cite{Otani:2010zza}, located on the beam axis, monitors the direction and stability of the neutrino beam. The ND280 detector~\cite{Amaudruz:2012agx,Assylbekov:2011sh,Abgrall:2010hi,Allan:2013ofa,Aoki:2012mf} is located at the same angle away from the beam axis as SK, and characterizes the rate of neutrino interactions from the beam before oscillations have occurred, thereby reducing systematic errors. ND280 is magnetized so that charged leptons and antileptons bend in opposite directions as they traverse the detector. This effect is used to measure the fraction of events in each beam mode that are from neutrino and antineutrino interactions. In this analysis, we select samples enriched in CCQE events and also several control samples enriched in interactions from other processes, allowing their rates to be measured separately. Here we use ND280 data that include a neutrino beam exposure of $5.8\times 10^{20}$ ($3.9\times 10^{20}$) protons hitting the T2K neutrino production target in neutrino(antineutrino)-mode. The explanation for the smaller data set in ND280 and its impact on the analysis method is described in the Methods Section.
 
\begin{figure}
  \centering
  \includegraphics[width=.49\textwidth]{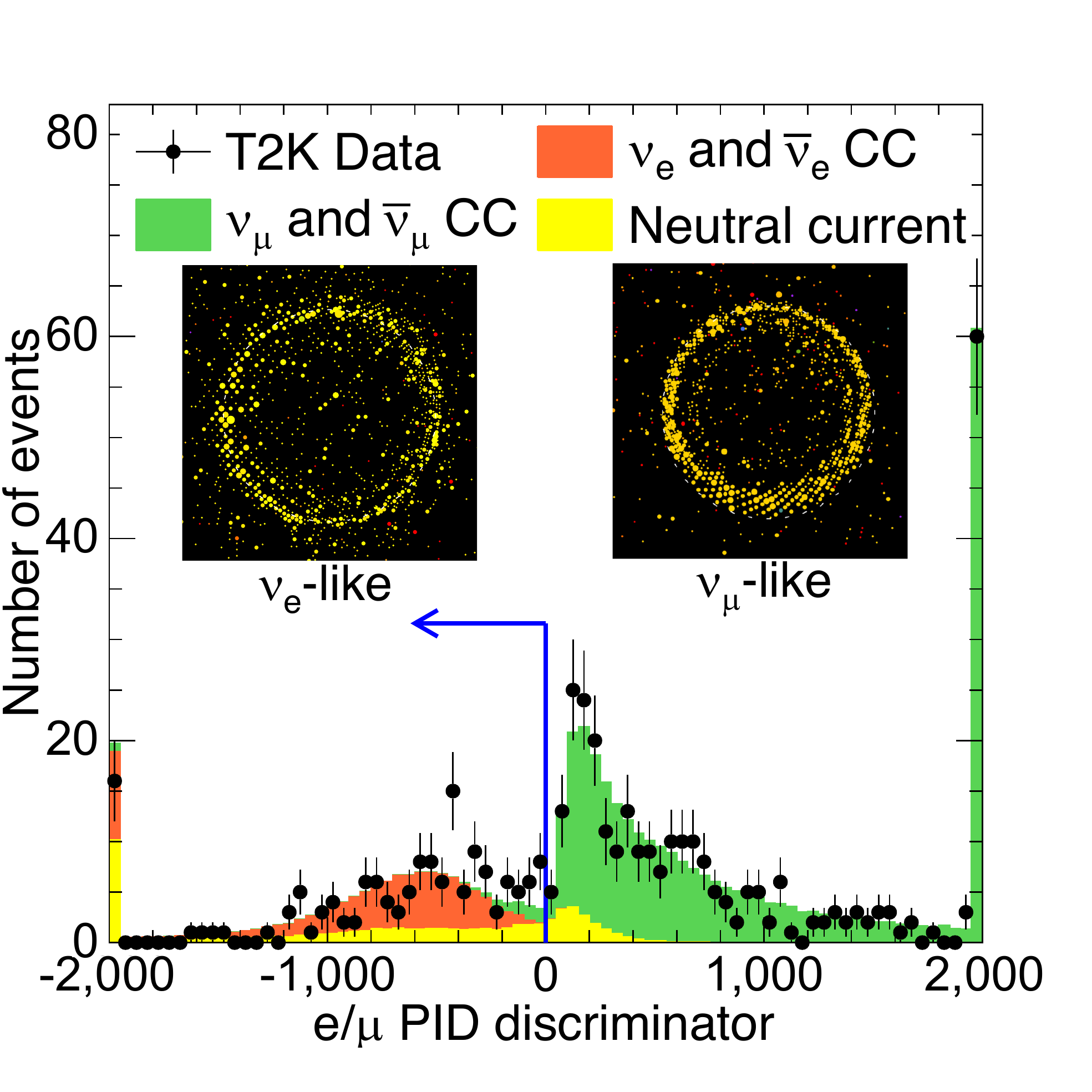}
  \caption{\textbf{Particle identification in the SK detector}. Distribution of the particle identification (PID) parameter used to classify Cherenkov rings as electron-like and muon-like. Events to the left of the blue line are classified as electron-like and those to the right as muon-like. The filled histograms show the expected number of single ring events after neutrino oscillations, with the first and last bins of the distribution containing events with discriminator values above and below the displayed range respectively. The vertical error bars on the data points are the standard deviation due to statistical uncertainty. The PID algorithm uses properties of the light distribution such as the blurriness of the Cherenkov ring to classify events. The insets show examples of an electron-like (left) and muon-like (right) Cherenkov ring.}
  \label{fig:skpid}
\end{figure}

SK is a 50~kt water detector instrumented with photo-multiplier tube  light sensors~\cite{Fukuda:2002uc}. In SK, Cherenkov light is produced as charged particles above a momentum threshold travel through the water. This light is emitted in ring patterns which are detected by the light sensors. Due to their lower mass, electrons scatter significantly more frequently (both elastically and inelastically) than muons so their Cherenkov rings are blurred. We use this blurring to identify the charged lepton's flavour, as illustrated in Figure~\ref{fig:skpid}. More information on the event reconstruction technique for SK data and the 
systematic uncertainty on SK modeling can be found in the Methods Section.
SK is not magnetized, so therefore relies on ND280's measurement of the neutrino and antineutrino composition of the beam in each mode.

We form five independent samples of SK events. For both neutrino- and antineutrino-beam mode there is a sample of events that contain a single muon-like ring (denoted 1$\mu$), and a sample of events that contain only a single electron-like ring (denoted 1$e$0de). These single-lepton samples are dominated by CCQE interactions. In neutrino-mode there is a sample containing an electron-like ring as well as the signature of an additional delayed electron from the decay of a charged pion and subsequent muon (denoted 1$e$1de). We do not use this sample in antineutrino-mode because charged pions from antineutrino interactions are mostly absorbed by a nucleus before they decay. Identifying both muon and electron neutrino interactions in both the neutrino- and antineutrino-mode beams allows us to measure the probabilities for four oscillation channels: $\nu_{\mu}\rightarrow\nu_{\mu}$ and $\bar\nu_{\mu}\rightarrow\bar\nu_{\mu}$, $\nu_{\mu}\rightarrow\nu_{e}$ and $\bar\nu_{\mu}\rightarrow\bar\nu_{e}$.

\begin{figure}
  \centering
  \includegraphics[width=.49\textwidth]{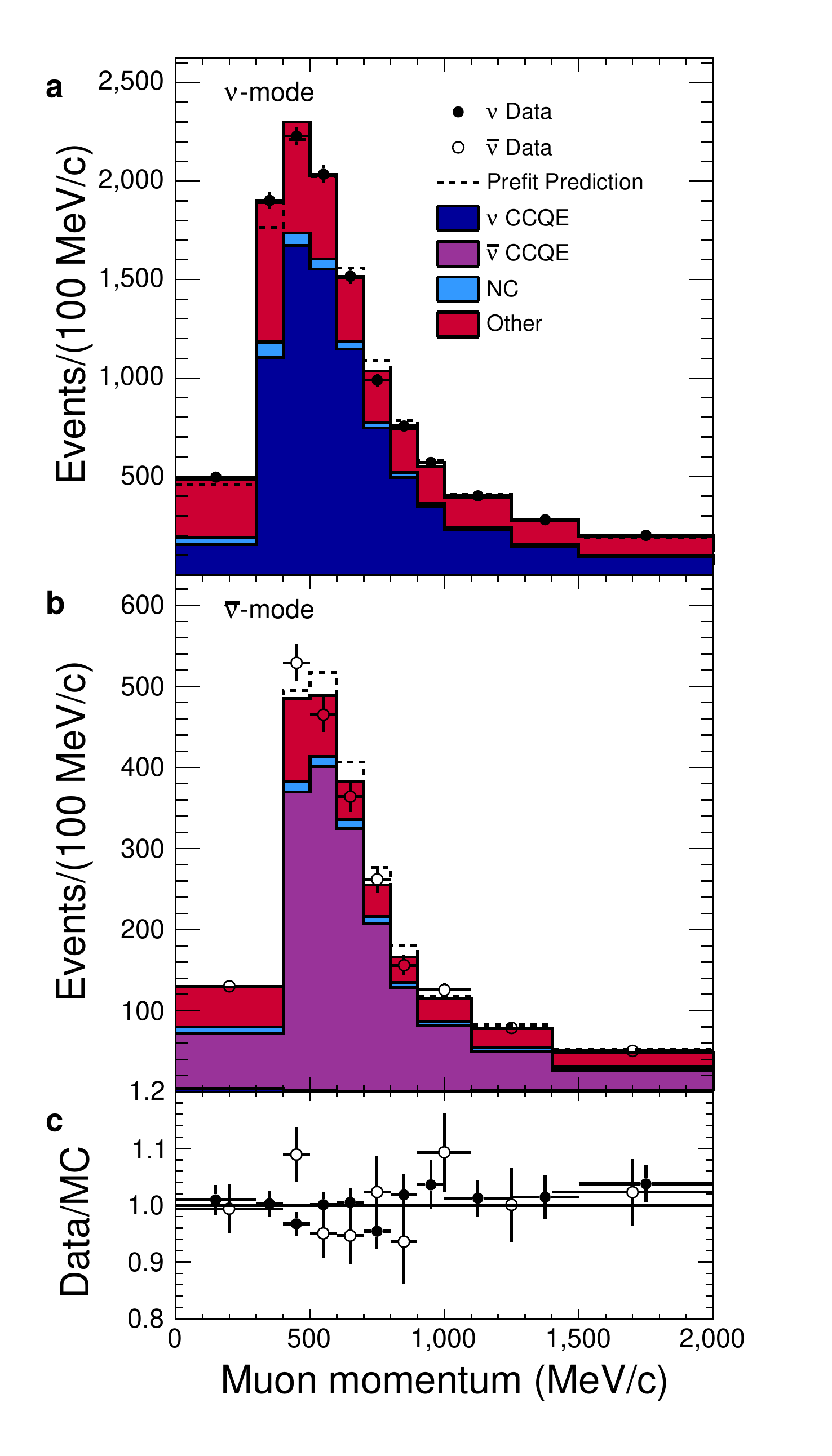}
  \caption{\textbf{Event predicition model tuning to near-detector data}. Reconstructed muon momentum in two of the ND280 CCQE-like event samples for both neutrino (subfigure \textbf{a}) and antineutrino (subfigure \textbf{b}) beam mode. The prediction with all parameters set to their best-fit value from a fit to the ND280 data is shown by the coloured histograms, split into true neutrino CCQE, antineutrino CCQE, neutral current (NC) and all other interactions. The dashed line shows the prediction before a fit to the ND280 data. The vertical error bars on the data represent the standard deviation due to statistical uncertainty. Subfigure \textbf{c} shows the ratio of the observed data to the best-fit prediction (MC) in both neutrino and antineutrino mode samples.}\label{fig:nd280_constraint}
\end{figure}

\begin{figure}
  \centering
  \includegraphics[width=.49\textwidth]{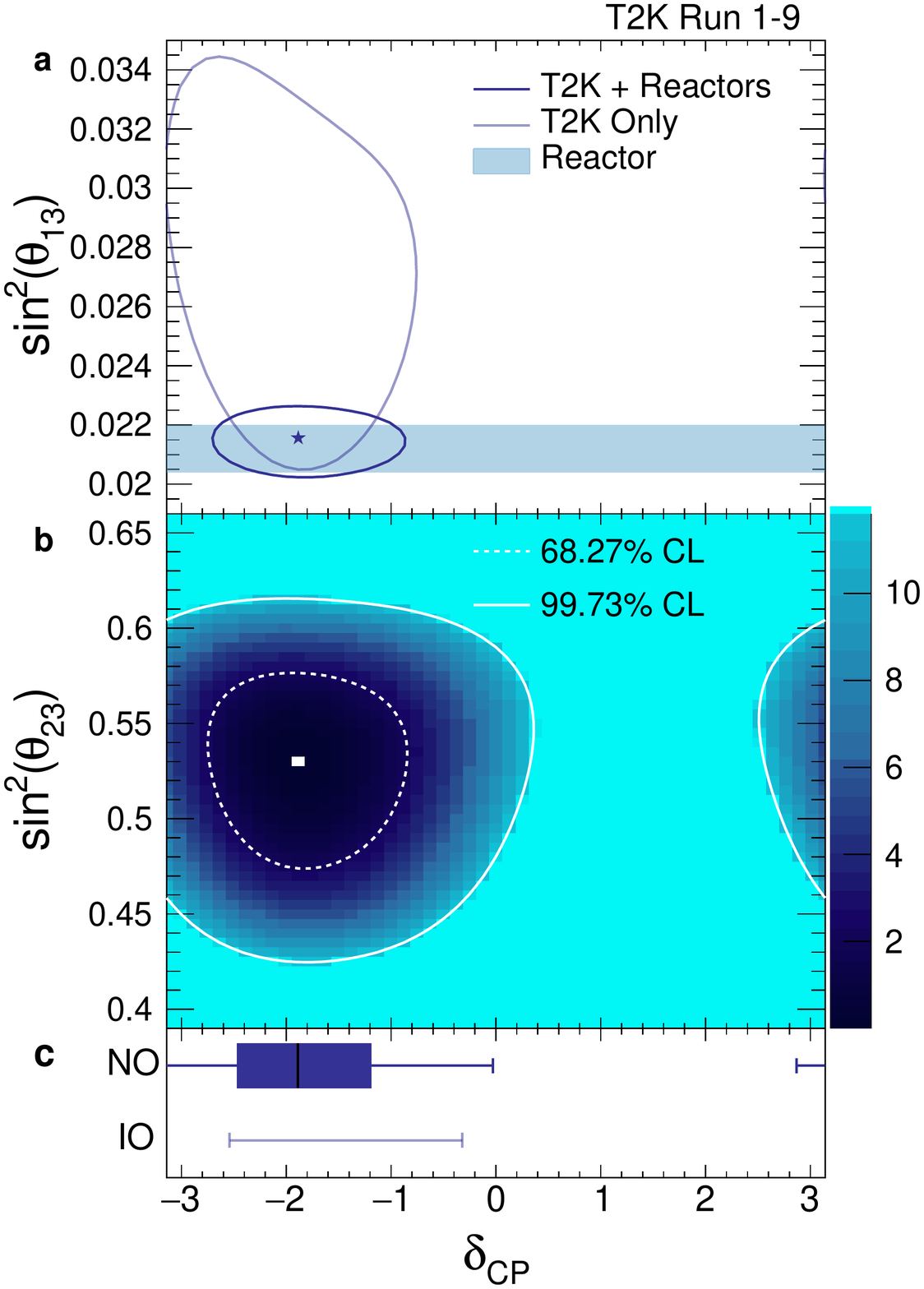}
  \caption{\textbf{Constraints on PMNS oscillation parameters}. Subfigure \textbf{a} shows 2D confidence intervals at the 68.27\% confidence level (CL) for $\delta_{CP}$ vs $\sin^2\theta_{13}$ in the preferred normal ordering. The intervals labelled T2K only indicate the measurement obtained without using the external constraint on $\sin^2\theta_{13}$, while the T2K + Reactor intervals do use the external constraint. The star shows the best-fit point of the T2K + Reactors fit in the preferred normal mass ordering. Subfigure \textbf{b} shows 2D confidence intervals at the 68.27\% and 99.73\% confidence level for $\delta_{CP}$ vs $\sin^2\theta_{23}$ from the T2K + Reactors fit in the normal ordering, with the colour scale representing the value of negative two times the logarithm of the likelihood for each parameter value. Subfigure \textbf{c} shows 1D confidence intervals on $\delta_{CP}$ from the T2K + Reactors fit in both the normal (NO) and inverted (IO) orderings. The vertical line in the shaded box shows the best-fit value of $\delta_{CP}$, the shaded box itself shows the 68.27\% confidence interval, and the error bar shows the 99.73\% confidence interval. It is notable that there are no values in the inverted ordering inside the 68.27\% interval.}\label{fig:dcp_theta13}
\end{figure}

We define a model of the expected number of neutrino events as a function of kinematic variables measured in our detectors with degrees of freedom for each of the oscillation parameters and for each source of systematic uncertainty. Systematic uncertainties arise in the modeling of neutrino-nucleus interactions
in the detector, the modeling of the neutrino production, and the modeling of the detector's response to neutrino interaction products.
Where possible, we constrain the model using external data. For example, the solar oscillation parameters, $\Delta m^{2}_{21}$ and $\sin^{2}(\theta_{12})$, whose values T2K is not sensitive to, are constrained using world average data~\cite{Tanabashi:2018oca}.  Whilst we are sensitive to $\sin^{2}\theta_{13}$, we use the combination of measurements from the Daya Bay, RENO and Double Chooz reactor experiments  to constrain this parameter ~\cite{Tanabashi:2018oca}, as they make a much more precise measurement than using T2K data alone (see Figure~\ref{fig:dcp_theta13}a). We measure the oscillation parameters by doing a marginal likelihood fit of this model to our near and far detector data. We perform several analyses using both Bayesian and frequentist statistical paradigms. Exclusive measurements of neutrino or antineutrino candidates in the near detector, one of which is shown in Figure~\ref{fig:nd280_constraint}, strongly constrain the neutrino production and interaction models, reducing the uncertainty on the predicted number of events in the four single-lepton SK samples from 13-17\% to 4-9\%, depending on the sample. The 1$e$1de sample's uncertainty is reduced from 22\% to 19\%.

A neutrino's oscillation probability depends on its energy, as shown in Eqs.~\ref{eq:nueprob} and ~\ref{eq:numuprob}. While the energy distribution of our neutrino beam is well understood, we cannot directly measure the energy of each incoming neutrino. Instead the neutrino's energy must be inferred from the momentum and direction of the charged lepton that results from the interaction. This inference relies on the correct modeling of the nuclear physics of neutrino-nucleus interactions. Modeling the strong nuclear force in multi-body problems at these energies is not computationally tractable, so approximate theories are used~\cite{Nieves:2005rq,Martini:2009uj,Benhar:2005dj,Salcedo:1987md}. The potential biases introduced by approximations in these theories constitute the largest sources of systematic uncertainties in this measurement. For scale, the largest individual source contributes 7.1\% of the overall 8.8\% systematic uncertainty on the single electron-like ring $\nu$-mode sample. Furthermore, as well as CCQE interactions, there are non-negligible contributions from interactions where additional particles were present in the final state but were not detected by our detectors. To check for bias from incorrect modeling of neutrino-nucleus interactions, we performed fits to simulated data sets generated assuming a range of different models of neutrino interactions \cite{Martini:2009uj,Benhar:2005dj}. We compared the measurements of the oscillation parameters obtained from these fits with the measurement from a fit to simulated data generated assuming our default model. We observed no significant biases in the obtained $\delta_{CP}$ best-fit values or changes in the interval sizes from any model tested. Biases are seen on $\Delta m^{2}_{32}$, and these have been incorporated in the analysis through an additional error of $3.9\times 10^{-5}$ $\text{eV}^{2}/c^{4}$ on the $\Delta m^{2}_{32}$ interval.  More details of the systematic uncertainties on neutrino interaction modeling can be found in the Methods Section.

The observed number of events at SK can be seen in Figure~\ref{fig:nueerecspectra}. The probability to observe an excess over prediction in one of our five samples at least as large as that seen in the electron-like charged pion sample is 6.9\% assuming the best-fit value of the oscillation parameters. We find the data shows a preference for the normal mass ordering with a posterior probability of 89\%, giving a Bayes factor of 8. We find $\sin^{2}(\theta_{23})= 0.53^{+0.03}_{-0.04}$ for both mass orderings. Assuming the normal (inverted) mass ordering we find $\Delta m^{2}_{32}= (2.45\pm 0.07)\times 10^{-3}$ $(\Delta m^{2}_{13}=(2.43\pm 0.07)\times 10^{-3})$ $\text{eV}^{2}/c^{4}$. For $\delta_{CP}$ our best-fit value and 68\% ($1\sigma$) uncertainties assuming the normal (inverted) mass ordering are $-1.89^{+0.70}_{-0.58} (-1.38^{+0.48}_{-0.54})$, with statistical uncertainty dominating. Our data show a preference for values of $\delta_{CP}$ which are near maximal $CP$ violation (see Figure \ref{fig:dcp_theta13}), while both $CP$ conserving points, $\delta_{CP}=0$ and $\delta_{CP}=\pi$, are ruled out at the 95\% confidence level. Here, we also produce 99.73\% ($3\sigma$) confidence and credible intervals on $\delta_{CP}$. In the favoured normal ordering the confidence interval contains [$-3.41$,$-0.03$] (excluding 46\% of the parameter space). We have investigated the effect of the excess seen in the 1$e$1de sample on this interval and find that had the observed number of events in this sample been as expected for the best-fit parameter values the interval would have contained [$-3.71$,$0.17$] ( excluding 38\% of parameter space). In the inverted ordering the confidence interval contains [$-2.54$,$-0.32$] (excluding 65\% of the parameter space). The 99.73\% credible interval marginalized across both mass orderings contains [$-3.48$,$0.13$] (excluding 42\% of the parameter space). The $CP$-conserving points are not both excluded at the 99.73\% level. However, this is the first time an experiment has reported closed 99.73\% ($3\sigma$) intervals on the $CP$-violating phase $\delta_{CP}$ (taking into account both mass orderings) and a large range of values around $+\pi/2$ are excluded. 

\begin{acknowledgments}
We thank the J-PARC staff for superb accelerator performance. We thank the 
CERN NA61/SHINE Collaboration for providing valuable particle production data. 
We acknowledge the support of MEXT, Japan; 
NSERC (Grant No. SAPPJ-2014-00031), NRC and CFI, Canada; 
CEA and CNRS/IN2P3, France; 
DFG, Germany; 
INFN, Italy; 
National Science Centre (NCN) and Ministry of Science and Higher Education, Poland; 
RSF (Grant \#19-12-00325) and Ministry of Science and Higher Education, Russia; 
MINECO and ERDF funds, Spain; 
SNSF and SERI, Switzerland; 
STFC, UK; and 
DOE, USA. 
We also thank CERN for the UA1/NOMAD magnet, 
DESY for the HERA-B magnet mover system, 
NII for SINET4, 
the WestGrid and SciNet consortia in Compute Canada, 
and GridPP in the United Kingdom. 
In addition, participation of individual researchers and institutions has been further supported by funds from ERC (FP7), 
"la Caixa” Foundation (ID 100010434, fellowship code LCF/BQ/IN17/11620050), the European Union’s Horizon 2020 Research 
and Innovation Programme under the Marie Sklodowska-Curie grant agreements no. 713673 and no. 754496, 
and H2020 Grants No. RISE-RISE-GA822070-JENNIFER2 2020 and RISE-GA872549-SK2HK; 
JSPS, Japan; 
Royal Society, UK; 
French ANR Grant No. ANR-19-CE31-0001; 
and the DOE Early Career program, USA.
\end{acknowledgments}

\clearpage

\section{Methods}
The measurement presented in this paper relies on the modeling of experimental apparatus to infer the parameters governing the oscillations of neutrinos.  This modeling can be broken into three main categories: the modeling of the neutrino production in the beam line, the modeling of neutrino-nucleus interactions in the detectors, and the modeling of the detectors' responses to final state particles and the inference of particle properties from the detector response.  The main sources of systematic uncertainty in the data analysis arise in these three areas, and
here we provide a description of the models and associated systematic uncertainties.

The inference of neutrino oscillation parameters from the data also relies on the statistical methods applied. In this section, we also provide a detailed description of the statistical methods used to infer the favored values and allowed regions for neutrino oscillation parameters.

\subsection{Neutrino Production Modeling}
The predicted neutrino and antineutrino fluxes, including the energies and flavours of neutrinos(antineutrinos), are estimated using a detailed simulation of the T2K beam line.  Measurement of the proton beam orbit, transverse width and divergence, and intensity are used as the initial conditions before simulating the interactions of protons in the T2K target to produce the secondary particles that decay to neutrinos.  Particle interactions and production inside the target are simulated with FLUKA2011~\cite{BOHLEN2014211,Ferrari:2005zk}, while particle propagation outside of the target is simulated with GEANT3~\cite{Brun:1994aa}. Interaction rates and hadron production in the simulation are tuned with hadron interaction data from external experiments, primarily the NA61/SHINE experiment which has collected data for T2K at the J-PARC proton beam energy of 30~GeV with the T2K target material of graphite~\cite{Abgrall:2015hmv}. Measurements of the magnetic horns' currents during beam operation and of the horns' magnetic fields before installation ensure accurate modeling of the charged particle focusing in the flux simulation.  The simulated fluxes are used as inputs to simulations of neutrino interactions and particle detection in the ND280 and SK detectors.  The spectrum of muon neutrinos (antineutrinos) produced from the decays of focused charged pions peaks at an energy of 0.6~GeV for an off-axis angle of 2.5$^{\circ}$.
Near the peak energy, 97.2\%\ (96.2\%) of the neutrino(antineutrino)-mode beam is initially $\nu_{\mu}$ $(\bar{\nu}_{\mu})$. The remaining components are mostly $\bar{\nu}_\mu$ ($\nu_\mu$); contaminations of $\nu_{e}+\bar{\nu}_{e}$ are only 0.47\% (0.49\%).

Since the neutrino flux prediction depends on the measured beam and beam line properties, which may vary with time, different flux predictions are made for each run period, and they are combined with weights proportional to the number of protons-on-target (POT) accumulated during periods of nominal detector operation.  The collected ND280 data corresponds to $5.8\times 10^{20}$ POT in neutrino mode and $3.9\times 10^{20}$ POT in antineutrino mode. This amount is less than the amount of data collected at SK due to lower efficiency of nominal data taking at ND280, and longer data processing times for ND280 data, limiting the available data set.  This POT difference between ND280 and SK is accounted for by combining the POT weighted flux predictions for each run period based on the beam exposures for the data collected at each detector.

The uncertainty on the flux calculation is evaluated by propagating uncertainties on the proton beam measurements, hadronic interactions, material modeling and alignment of beam line elements, and horn current and field measurements.  In each case, variations of the source of uncertainty are considered and the effect on the flux simulation is evaluated.  The INGRID on-axis neutrino detector is not used to tune the beam direction during operation.  Hence, it provides an independent measurement of the beam direction~\cite{Abe:2011xv}, which is used to validate the flux simulation.  The uncertainty on the INGRID beam direction measurement is propagated in the flux model. The variations are used to calculate covariances for the flux prediction in bins of energy, flavour, neutrino/antineutrino mode and detector (ND280 and SK).  These covariances are used to propagate uncertainties on the flux prediction in the oscillation analysis. The dominant source of systematic uncertainty is from the hadron interaction data and models.  The uncertainty on the flux normalization in this analysis near the peak energy of 0.6~GeV is 9\%. In future analyses we aim to improve this to approximately 5\% by using NA61/SHINE particle production data measured from a replica of the T2K target~\cite{Abgrall:2016jif,Berns:2018tap}. Uncertainties on the proton beam orbit and alignment of beam line elements correspond to an uncertainty on the off-axis angle at the ND280 and SK detectors, corresponding to uncertainties on the peak energy of the neutrino spectrum at those detectors.

\subsection{Neutrino Interaction Modeling}
The T2K detectors measure products of neutrinos and antineutrinos interacting on nuclei and free protons with energies ranging from $\sim$0.1~GeV to 30~GeV.  These interactions are modeled with the NEUT~\cite{NEUT} neutrino interaction generator, using version 5.3.2.  NEUT uses a range of models to describe the physics of the initial nuclear state, the neutrino-nucleon(s) interaction, and the interactions of final state particles in the nuclear medium.  

The primary signal processes in SK are defined by the presence of a single charged lepton candidate with no other visible particles.  The dominant process at the peak energy of 0.6~GeV is Charged-Current Quasi-Elastic (CCQE) scattering.  This process corresponds to the neutrino or antineutrino scattering on a single nucleon bound in the target nucleus.  The neutrino-nucleon scattering in NEUT is implemented in the formalism of Llewellyn-Smith~\cite{LlewellynSmith:1971uhs}.  For the initial nuclear state, NEUT implements a relativistic Fermi gas (RFG) model of the target nucleus, including long-range correlations evaluated using the random phase approximation (RPA)~\cite{nieves_rpa,*nieves_rpa_erratum}.  NEUT includes an alternative initial state model based on spectral functions describing the initial momentum and removal energy for bound nucleons~\cite{Benhar-SF}.  

Additional processes that can produce a signal-like final state are modeled in NEUT. The 2p-2h model of Nieves {\em et al.}~\cite{nieves,nievesExtension} predicts production of multinucleon excitations, where more than one nucleon and no pions are ejected in the final state.  The ejected nucleons are typically below detection threshold in a water Cherenkov detector, making this process indistinguishable from the CCQE process.

The signal candidate sample with one prompt electron-like ring and the presence of an electron from muon decay consists primarily of interactions where a pion is produced.  These single-pion interactions can also populate the samples without an additional electron from muon decay if the pion is absorbed in the target nucleus or on a nucleus in the detector, or if it is not detected. Processes producing a single pion and one nucleon are described by the Rein-Sehgal model~\cite{ReinSehgal}. Processes with multiple pions are simulated with a custom model below 2~GeV of hadronic invariant mass and by PYTHIA~\cite{Sjostrand:1993yb} otherwise. These processes may be selected as events with single Cherenkov rings if the pion is absorbed in the target nucleus or surrounding nuclei, or if it is not detected. The final state interactions of pions and protons in the target nucleus are modeled with the NEUT intranuclear cascade model where the density dependence of the mean free path for pions in the target nucleus is calculated based on the $\Delta$-hole model of Oset {\it et. al.}~\cite{Oset:1986sy} at low momenta and from p-$\pi$ scattering data from the SAID database at high momenta.  The microscopic interaction rates for exclusive pion scattering modes are then tuned to macroscopic $\pi$-nucleus scattering data. 

We consider two types of systematic uncertainties on neutrino interaction modeling in the oscillation measurement.  In the first, parameters in the nominal interaction model are allowed to vary and are constrained by ND280 data.  These parameters are then marginalized over when measuring oscillation parameters.  They include uncertainties on nucleon form factors, the corrections for long-range correlations, the rates of different neutrino interaction processes, the final state kinematics of the CCQE, 2p-2h and single pion production processes, and the rates of pion final state interactions.  Most of these are parameters in the models with physical interpretations, and they modify the overall rate of interactions, the final state topology, and the kinematics of final state particles. After the constraint from ND280 data, these parameters are correlated with the systematic parameters in the neutrino production model, and their combined impact on the predicted event distributions in Super-K is evaluated.  The constrained interaction model and neutrino production model parameters contribute a 2.7\% uncertainty on the prediction of the relative number of electron neutrino and electron antineutrino candidates, the third largest source of systematic uncertainty, as shown in Supplementary Table I.

We also include an uncertainty on the $\nu_{e}$ and $\bar{\nu}_{e}$ cross sections relative to the $\nu_{\mu}$ and $\bar{\nu}_{\mu}$ cross sections. This introduces a direct uncertainty on the relative prediction of $\nu_{e}$ and $\bar{\nu}_{e}$ candidates, and is motivated by uncertainties in the neutrino-nucleon scattering cross section arising from the charged lepton masses~\cite{McFarland-Day}.  As shown 
in Supplementary Table I, this introduces a relative uncertainty of 3.0\%, the second largest single source of systematic uncertainty
in the $CP$ asymmetry measurement.

The second type of systematic uncertainty is evaluated by introducing simulated data generated with an alternative model into the analysis and evaluating the impact on measured oscillation parameters. This approach is used to evaluate the effect of changing the nuclear initial state model including the use of the spectral functions and changes to the removal energy for initial state nucleons. This approach is also applied to evaluate the impact of changes to the 2p-2h interaction cross section as a function of energy, using an alternative single pion production model~\cite{Rein:1987cb,Kabirnezhad:2017jmf}, and applying alternative multi-pion production tuning~\cite{Yang:2009zx}. The largest biases observed in this approach are on the measurement of $\Delta m^{2}_{32}$, while the impact on other parameters is typically small compared to the total systematic uncertainty.  In the case of $\Delta m^{2}_{32}$, an additional uncertainty of $3.9\times10^{-5}$~eV$^{2}$/c$^{4}$ is included by taking a convolution of a Gaussian of width $3.9\times10^{-5}$~eV$^{2}$/c$^{4}$ with the likelihood. For the measurement of the other oscillations parameters, it was found that the biases introduced by varying the removal energy by up to 18~MeV for initial state nucleons were not negligible. An additional uncertainty equal to the bias in the predicted event distributions when varying the removal energy by 18~MeV was added to the analysis. As shown in Supplementary Table I, this introduces a 3.7\% uncertainty on the relative prediction of electron neutrino and electron antineutrino candidates, the largest single source of systematic uncertainty in the analysis.

\subsection{Super-Kamiokande Event Reconstruction}
Photosensors installed on the SK inner detector register Cherenkov light produced as charged particles produced by neutrino interactions travel through the water volume \cite{Fukuda:2002uc}. Photosensor activity clustered in time, on the order of a micro-second, is called an event. Events coincident with the T2K-beam timing are selected as candidate beam neutrino interactions.

Neutrino interaction events in SK often have multiple periods of photosensor activity separated in time within an event. The most frequent example is a muon decaying into an electron. A decay electron can be used to tag a muon even when the muon energy is below the Cherenkov threshold, e.g. the case that the muon is produced by a charged pion decaying at rest. Such sub-events are searched for with a peak finding algorithm and reconstructed separately in later processes.

The kinematics of the charged particles are reconstructed from the timing and the number of detected photons of each photosensor signal by using a maximum likelihood algorithm~\cite{10.1093/ptep/ptz015}. The likelihood consists of the probability of each photosensor to detect photons or not and the charge and timing probability density functions of the hit photosensors. This new reconstruction algorithm makes use of the timing and charge information obtained by all the photosensors simultaneously, which leads to better kinematic resolutions and particle classifying performances compared to the previously used reconstruction algorithm.

The five signal samples are formed by using the reconstructed event kinematics. All the selected events are required to have little photosensor activity in an outer veto detector, and the reconstructed neutrino interaction position is required to be inside the inner detector fiducial volume. The reconstruction improvement enabled us to extend the fiducial volumes used in the analysis.
We performed a dedicated study to optimize the fiducial volume to maximize T2K sensitivities to oscillation parameters taking into account both the statistical and systematical uncertainties. The position dependent SK detector systematics are estimated by using SK atmospheric neutrino interaction events. The fiducial volume expansion contributes to the increase of selected electron-like (muon-like) events by 25\% (14\%)~\cite{Abe:2018wpn}.

Systematic uncertainties regarding SK detector modeling were addressed by various control samples. Uncertainties on the position reconstruction bias and on the decay electron tagging are estimated by using cosmic-ray muons stopping inside the inner detector. Simulated atmospheric neutrino events are compared to data to evaluate systematic uncertainties on the modeling of signal selection efficiencies and the background contamination of the five analysis samples. Parameters describing possible mis-modeling of Cherenkov ring counting and particle identifications are introduced and constrained by a fit to the control samples.  These parameters are varied according to the posterior distribution from the fit to the control samples, and the uncertainties on the T2K samples of interest are evaluated. The uncertainty on the modeling of the efficiency to select events with neutral pions, which is one of the dominant backgrounds in the electron neutrino CCQE-like event sample, is estimated by constructing a set of hybrid events that combine one data and one simulated electron-like Cherenkov ring to imitate the decay of a neutral pion.
The uncertainties on the numbers of selected total events in SK are 2--4\% depending on the signal categories.  As shown in Supplementary Table I, the relative uncertainty on the predicted number of electron neutrino and electron antineutrino candidates for samples with no decay electrons is 1.5\%.

\subsection{Statistical Methods}
We use a binned likelihood-ratio method comparing the observed and predicted numbers of muon- and electron-neutrino candidate events in our five samples. In neutrino beam mode these are electron-like, muon-like and electron-like charged pion samples, while in antineutrino beam mode these are electron-like and muon-like samples. The samples are binned in reconstructed energy and, for the electron-neutrino-like samples, the angle between the lepton and the beam direction. In particular, best-fits are determined by minimising the sum of the following likelihood function (marginalized over nuisance parameters) over all of our samples
\begin{equation}\label{eq:likelihood}
    \begin{split}
	-2\ln\lambda(\overline{\delta_{CP}};\mathbf{a}) = 2\sum_{i=1}^{N}\left[ n_i^{obs}\ln\left( \frac{n_i^{obs}}{n_i^{exp}} \right) + n_i^{exp} - n_i^{obs} \right]\\
	+(\mathbf{a} - \mathbf{a_0})^T \mathbf{C^{-1}}(\mathbf{a} - \mathbf{a_0})
    \end{split}
\end{equation}
where $\overline{\delta_{CP}}$ is the estimated value of $\delta_{CP}$, $\mathbf{a}$ is the vector of systematic parameter values (including the remaining oscillation parameters), $\mathbf{a_0}$ is the vector of default values of the systematic parameters, $\mathbf{C}$ is the systematic parameter covariance matrix, $N$ is the number of reconstructed energy and lepton angle bins, $n_i^{obs}$ is the number of events observed in bin $i$ and $n_i^{exp} = n_i^{exp}(\overline{\delta_{CP}};\mathbf{a})$ is the corresponding expected number of events. Systematic parameters are marginalized according to their prior constraints from the fit to ND280 data.

We perform both frequentist and Bayesian analyses of our data. The measurement of $\delta_{CP}$ from each of the analyses is in agreement, with the presented confidence intervals coming from a frequentist analysis and the Bayes factors and credible intervals coming from a Bayesian analysis. In the frequentist analysis a fit is first performed to the near detector samples binned in the momentum and cosine of the angle between the lepton and the beam direction, with penalty terms for flux, cross-section and detector systematic parameters at the near detector. Systematic parameter constraints are then propagated from the near to the far detector via the covariance matrix, $\mathbf{C}$, in Eq.~\ref{eq:likelihood} and their fitted values. The matrix is the combination of the posterior covariance from the near detector fit with the priors for the oscillation parameters, with some parameters affecting both detectors directly, while others that affect only the far detector are constrained through their correlation with near detector affecting parameters. Gaussian priors for $\sin^2(\theta_{13})$, $\sin^2(\theta_{12})$, and $\Delta{}m^2_{21}$ are taken from the Particle Data Group's (PDG) world combinations~\cite{Tanabashi:2018oca}, while $\sin^2(\theta_{23})$ and $\Delta{}m^2_{32}$ ($\Delta{}m^2_{13}$) have uniform priors in normal (inverted) mass ordering. For the Bayesian analyses the prior for $\delta_{CP}$ is uniform, with an additional check applying a uniform prior in $\sin(\delta_{CP})$ producing the same conclusions. Furthermore, rather than fitting the near detector and propagating to the far detector as a two step process, the Bayesian analysis directly includes the near detector samples in its expression for the likelihood and therefore performs a simultaneous fit of the near and far detector data.

The neutrino oscillation probability depends non-linearly on the oscillation parameters, with different possible values of $\delta_{CP}$ corresponding to a bounded enhancement or suppression of the electron (anti)neutrino appearance probability. If statistical fluctuations in the data exceed these bounds they are not accommodated by the model, and as a result the critical $\Delta\chi^2$ value for a given confidence level is often different from the asymptotic rule of Wilks \cite{Wilks1938}. To address this problem the frequentist analysis constructs Neyman confidence intervals using the approach described by Feldman and Cousins~\cite{FC} and thus critical values of $\Delta\chi^2$ vary as a function of $\delta_{CP}$ and the mass ordering. The critical values at a given confidence level are determined by fitting at least 20,000 simulated datasets for each given true value of $\delta_{CP}$ and the mass ordering. The remaining oscillation parameters are varied according to their priors. In particular, for $\sin^2(\theta_{13})$, $\sin^2(\theta_{12})$, and $\Delta{}m^2_{21}$ these priors are taken from the PDG~\cite{Tanabashi:2018oca}, with $\sin^2(\theta_{13})$ determined by the reactor experiments noted in the main text. For $\sin^2(\theta_{23})$, and $\Delta{}m^2_{32}$ ($\Delta{}m^2_{13}$) the priors take the form of likelihood surfaces produced from fits of simulated datasets. The simulated datasets are generated using oscillation parameter best-fits in normal and inverted mass orderings. The remaining systematic parameters are varied according to their prior constraints from the fit to ND280 data.

The Bayesian analysis uses Markov Chain Monte Carlo (MCMC) to take random samples from the likelihood. The particular MCMC algorithm used is Metropolis-Hastings~\cite{Hastings1970}. For a sufficiently large number of samples the Markov chain achieves an equilibrium probability distribution. The number of steps in the chain with a particular value of a parameter is proportional to the posterior probability for the parameter to have that value marginalized over all the other parameters. Credible intervals are then formed on the basis of highest posterior density, with bins of equal width in the parameter under study. Given the arbitrary initial state of the Markov chain, a finite number of samples must be obtained to allow the chain to converge to a state in which it is correctly sampling from the distribution. These preliminary `burn-in' samples are discarded.

\section{Author Contributions}
The operation, Monte Carlo simulation, and data analysis of the T2K Experiment are carried out by the T2K Collaboration with contributions from all collaborators listed as authors on this manuscript.  The scientific results presented here have been presented to and discussed by the full collaboration, and all authors have approved the final version of the manuscript.

\section{Data Availability}
The likelihood surface data that support these findings will be made available for public access on http://t2k-experiment.org.

\section{Code Availability}
The T2K collaboration develops and maintains the code used for the simulation of the experimental apparatus and statistical analysis of the raw data used in this result.  This code is shared among the collaboration, but not publicly distributed.  Inquiries regarding the algorithms and methods used in this result may be directed to the corresponding author.

\thispagestyle{empty}

\begin{table*}[ht]
\sffamily
{\sffamily Supplementary Table 1: The systematic uncertainty on the predicted relative number of electron neutrino and electron antineutrino
candidates in the Super-K samples with no decay electrons.}
        \centering
        \begin{tabular}{l|c}
                \hline
                Type of Uncertainty  & $\nu_{e}/\bar{\nu}_{e}$ Candidate Relative Uncertainty (\%) \\ \hline
                Super-K Detector Model & 1.5 \\ 
                Pion Final State Interaction and Rescattering Model & 1.6 \\ 
                Neutrino Production and Interaction Model Constrained by ND280 Data & 2.7 \\ 
                Electron Neutrino and Antineutrino Interaction Model & 3.0 \\ 
                Nucleon Removal Energy in Interaction Model & 3.7 \\ 
                Modeling of Neutral Current Interactions with Single $\gamma$ Production & 1.5 \\ 
                Modeling of Other Neutral Current Interactions & 0.2 \\ \hline 
                Total Systematic Uncertainty & 6.0 \\ \hline
        \end{tabular}
\end{table*}

\end{document}